\documentclass[10pt,aps,prd,nofootinbib,reprint,groupedaddress]{revtex4-1}

\usepackage[utf8]{inputenc}
\usepackage{amsmath}
\usepackage{graphicx}
\usepackage{amssymb}

\usepackage{hyperref}
\hypersetup{colorlinks=true,allcolors=[rgb]{0.5,0.0,0.5}}

\newcommand{\orcid}[1]{\href{https://orcid.org/#1}{#1}}

\newcounter{RSQ}

\newcommand{\lsim}{\lesssim}
\newcommand{\gsim}{\gtrsim}
\newcommand{\eq}[1]{Eq.~(\ref{#1})}
\newcommand{\ord}[1]{\mathcal{O}{(#1)}}
\newcommand{\beq}{\begin{equation}}
\newcommand{\eeq}{\end{equation}}
\newcommand{\bea}{\begin{eqnarray}}
\newcommand{\eea}{\end{eqnarray}}

\newcommand{\vphi}{\varphi}

\usepackage[normalem]{ulem}

\newcommand{\appropto}{\mathrel{\vcenter{
  \offinterlineskip\halign{\hfil$##$\cr
    \propto\cr\noalign{\kern2pt}\sim\cr\noalign{\kern-2pt}}}}}

\begin{document}

\pagestyle{plain}

\title{Muon \boldmath$g-2$ and a Geocentric New Field}

\author{Hooman Davoudiasl}
\email{hooman@bnl.gov}
\thanks{\orcid{0000-0003-3484-911X}}

\author{Robert Szafron}
\email{rszafron@bnl.gov}
\thanks{\orcid{0000-0002-9640-6923}}

\affiliation{High Energy Theory Group, Physics Department, Brookhaven National Laboratory, Upton, NY 11973, USA}

\date{\today}


\begin{abstract}

Light scalars can in principle couple to both bulk matter and fermion spin, with  hierarchically disparate strengths.  Storage ring measurements of fermion electromagnetic moments via spin precession can be sensitive to such a force, sourced by the Earth.  We discuss how this force could lead to a deviation of the measured muon anomalous magnetic moment, $g-2$, from the Standard Model prediction.  Due to its different parameters, the proposed JPARC muon $g-2$ experiment can provide a direct test of our hypothesis.  A future search for the proton electric dipole moment can have good sensitivity for the coupling of the assumed scalar to nucleon spin.  We also argue that supernova constraints on axion-muon coupling may not be applicable in our framework.

\end{abstract}
\maketitle



The Standard Model (SM) of particle physics, though successful at precisely describing a great many observed phenomena, leaves a number of important questions unanswered.  Among these, the unknown nature of cosmic dark matter is perhaps the starkest manifestation of the need for extending the SM.  Also, a variety of conceptual questions -- for example regarding a   quantum formulation of gravity -- provide additional strong motivation for invoking new physics.  It is then reasonable to suspect that there could be new long range forces, originating in  sectors beyond the SM,  that feebly interact with ordinary matter.  Needless to say, uncovering any such interaction would revolutionize our understanding of the Universe and open up new fronts in fundamental physics.  Therefore, this is a possibility that is worth investigating in both theory and experiment. 

In this {\it Letter}, motivated by the above considerations, we examine  a possible long range force mediated by a light scalar $\vphi$ with a  derivative coupling to one or more SM fermions; this interaction respects a shift symmetry, much like a conventional Goldstone boson.  In addition, we entertain the possibility that the shift symmetry is very weakly broken by a Yukawa coupling to ordinary matter, which could be electrons or nucleons.  We further assume that the scalar mass $m_\vphi < R_\oplus^{-1}$, where $R_\oplus \approx 6368$~km $\approx (3.1\times 10^{-14}~\text{eV})^{-1}$ is the mean radius of the Earth. Hence, the Earth can act as a coherent source for $\vphi$, which leads to a field that falls like $1/r^2$ with distance $r \geq R_\oplus$ from the center of the Earth.  This field, which represents the radial  variation  $\partial_r\vphi$ of the scalar, can then couple to the spin of a fermion and make it precess, similar to the effect of a feeble magnetic field pointing towards or away from the center of the Earth.  

Precision measurements at current and proposed storage rings can in principle be sensitive to the aforementioned anomalous precession caused by a terrestrial $\vphi$ field.  In particular, the longstanding apparent deviation of the muon $g-2$ from SM prediction \cite{Muong-2:2006rrc,Aoyama:2020ynm} 
can potentially be explained in this scenario.  This hypothesis can also be tested at a future proton storage ring facility if the scalar also has derivative coupling to nucleons. In what follows, we will present a simple underlying model and relevant numerical estimates, in order to elucidate the above points.  Probes of light scalars at storage ring facilities, in contexts different from the one examined here, have also been discussed in Ref.~\cite{Graham:2020kai} and  Ref.~\cite{Janish:2020knz}; the latter considers a possible relation to the muon $g-2$ anomaly.    A similar model, in a different regime of parameters, was introduced in  Ref.~\cite{Kim:2021eye}, where the consequences of the CP violating and CP preserving couplings of an  ultralight pseudo-scalar to Galactic dark matter and the SM fermions, respectively, were studied.   

{\it A model.---}  We will assume the following interactions for $\vphi$
\beq
g_s^f \vphi \bar f f + \frac{g_a^\psi}{M}\,\partial^\mu \vphi\, \bar\psi \gamma_\mu \gamma_5 \psi\,,
\label{int}
\eeq
where $g_s^f$ and $g_a^\psi$ are particle species-dependent constants and $M$ sets the scale of the dimension-5 operator.  For the purposes of this work, $f$ could be a nucleon $N$ or an electron, so that the Earth could source a significant $\vphi$ field.  Depending on the relevant experimental probe, $\psi$ is one of a few known fermions; here we will mostly consider the muon $\mu$ or the proton $p$.

Let us consider the case $f=N$.  Then, there will be a geocentric background $\vphi$ field for $r$ larger than $R_\oplus$, but much smaller than the Earth-Moon distance, given by 
\beq
\vphi(r) \approx \frac{g_s^N\,{\cal N}_\oplus}{4 \pi\,  r}\,,
\label{vphi}
\eeq
where the number of nucleons in the Earth is given by ${\cal N}_\oplus\approx 3.6\times 10^{51}$ \cite{ParticleDataGroup:2020ssz}.  The above field may receive contributions from other bodies, but if we assume that the range of $\vphi$ is not much larger than the size of the Earth, we can ignore such corrections.  In that case, to a very good approximation, $\vphi$ is static and hence  $\partial_t \vphi=0$. The interactions in \eq{int} then yield a coupling between the scalar and the spin $\vec \sigma$ of $\psi$ of the form $\vec{\nabla} \vphi.\vec{\sigma}$.  Following the discussion in Ref.~\cite{Graham:2020kai}, we find that,  in our setup, this interaction would lead to a laboratory frame  angular precession frequency   $\omega_\vphi^\oplus$ at the surface of the Earth, given by 
\beq
\omega_\vphi^\oplus \approx 
\frac{g_s^N g_a^\psi{\cal N}_\oplus}{\gamma_\psi M \,R_\oplus^2}\,,
\label{omegaE}
\eeq
where $\gamma_\psi$ is the time dilation factor, due to the motion of $\psi$. 

{\it Contribution to the muon $g-2$}.---Let us examine whether the above precession can account for the discrepancy between muon $g-2$ theory and experiment, corresponding to $\psi=\mu$.  To do so, we need to determine the allowed values for the parameters entering \eq{omegaE}.  Using the latest results from the MICROSCOPE collaboration \cite{MICROSCOPE:2022doy}, in conjunction with the analysis in Ref.~\cite{Fayet:2017pdp}, we find the 2 $\sigma$ bound 
$|g^N_s|<8.0\times 10^{-25}$.  

The analysis in Ref.~\cite{Bollig:2020xdr}, based on supernova SN1987A limits, yields the constraint $|g_a^\mu| < 4.0 \times 10^{-8}$ for the coupling of $\vphi$ to $\mu$, where we will set $M=1$~GeV henceforth (the corresponding results of  Ref.~\cite{Croon:2020lrf} are somewhat stronger, but the corrected limit obtained in Ref.~\cite{Bollig:2020xdr} represents the latest update of the relevant calculation).  However, we note that our assumption of Yukawa interactions for $\vphi$ could nullify this constraint, as explained in the Appendix.  We note that the constraint from Cosmic Microwave Background is subject to uncertainties, but could be at the level of $g_a^\mu \lsim 10^{-4}$ \cite{DEramo:2018vss,Bollig:2020xdr}.    However, this constraint implicitly assumes a reheat temperature $T_{RH}\gsim m_\mu$, which is not a necessary cosmological assumption; in fact $T_{RH}$ can be as low as $\sim 5$~MeV \cite{Hannestad:2004px}.  Based on the above considerations, we conservatively assume $g_a^\mu \lsim 3.0\times 10^{-5}$.  

The Brookhaven (E821) \cite{Muong-2:2006rrc} and Fermilab (E989)  \cite{Muong-2:2021ojo} storage ring measurements of the muon $g-2$ are conducted at the ``magic momentum" corresponding to $\gamma_\mu = 29.3$.  To get an estimate of the effect from the coupling of $\vphi$ to muons, we take  $g_s^N g_a^\mu < 2.4 \times 10^{-29}$, which is consistent with the above bounds.
We thus obtain    
\beq
\omega_\vphi^\oplus < 4.3 ~\text{rad/s}\,.
\label{omega_val}
\eeq
The value measured by experiment corresponds to 
$\omega_a\approx 1.4 \times 10^6$~rad/s, which yields 
\beq
\frac{\omega_\vphi^\oplus}
{\omega_a} < 3.0\times 10^{-6}.
\label{omega-ratio}
\eeq

Note that since $\vec{\nabla}\vphi$ is a radial field, it is aligned, to an excellent approximation, with the magnetic field of the muon storage ring \cite{Muong-2:2006rrc,Muong-2:2021ojo}. The Brookhaven experiment E821 \cite{Muong-2:2006rrc} measured $a_\mu$ for $\mu^+$ and $\mu^-$, which required flipping the direction of the magnetic field in order to store $\mu^-$.     This flip compensates the opposite signs of $\mu^+$ and $\mu^-$ in the interaction term of spin with the  magnetic field in the Hamiltonian. The effect of $\vec{\nabla}\vphi$ is the same for particles and antiparticles, thus both for muons and antimuons the change in the precession frequency induced by $\vphi$ is the same.  Note that the $B$ field at the Fermilab experiment points in the same way for $\mu^+$ storage as the Brookhaven experiment \cite{Muong-2:2021vma}.  Thus, $\omega_\vphi^\oplus$ can be simply added or subtracted from $\omega_a$, that is $\omega_\vphi^\oplus \approx \delta \omega_a$.  
Since the measured value of the muon $(g-2)/2$, denoted by $a_\mu$, is proportional to $\omega_a$ it follows that 
\beq
\frac{\delta a_\mu}{a_\mu}  \approx \frac{\omega_\vphi^\oplus}{\omega_a}.
\label{delta-amu}
\eeq
Hence, we find 
\beq
\delta a_\mu < 3.5\times 10^{-9}, 
\label{delamu}
\eeq
from \eq{omega-ratio}, where $a_\mu \approx \alpha/(2\pi)$ \cite{Schwinger:1948iu}, and $\alpha\approx 1/137$.  We thus see that the effective inferred $\delta a_\mu$ above can accommodate the current discrepancy \cite{Muong-2:2021ojo}
\beq
\Delta a_\mu = (251 \pm 59)\times 10^{-11} 
\label{currentg-2}
\eeq
between the measured and predicted values of $a_\mu$.  However, we note that the status of theory is still under investigation  \cite{Borsanyi:2020mff} and the value of the deviation may change as calculations get more refined.

{\it Further probes.}---
The planned muon $g-2$/EDM experiment at J-PARC \cite{Abe:2019thb} allows us to test our scenario, directly.  With the $\mu^+$ boost $\gamma_\mu\approx 3$ approximately ten times smaller, and a magnetic field $B\approx 3$~T about twice as large as the  magnetic field in the Brookhaven/Fermilab storage ring, the deviation from the SM value of $a_\mu$ induced by $\vphi$ field should be about five times larger.  Note that the current design for the experiment at JPARC has the $B$ field pointing the same way as the Brookhaven/Fermilab field for $\mu^+$ storage \cite{Abe:2019thb}.

Here, we note that if the proton spin also couples to $\vec\nabla \vphi$, corresponding to $\psi=p$, one may probe this interaction in a storage ring proposed for detecting proton EDM at the level of $10^{-29}~e$.cm \cite{Omarov:2020kws,Alexander:2022rmq}.  The magic momentum used in order to achieve a frozen spin condition for the proton is $0.7$~GeV \cite{Omarov:2020kws}, corresponding to a small boost $\gamma_p\approx 1.25$.    This proposal employs counter-rotating proton beams in order to suppress systematic uncertainties.  The presence of $\vec\nabla \vphi$ in our model will show up as a spurious vertical $B$ field leading to precession of the longitudinally polarized protons in the plane of storage ring.  This apparent source of error  would need to be corrected by applying a compensatory  vertical $B$ field, which would lead to spatial separation of counter-rotating beams.  The experimental tolerance for this effect is $\sim 10^{-3}$~rad/s.  For the currently allowed parameters $g_s^N \sim 8\times 10^{-25}$, and the proton coupling $g_a^p \sim   10^{-9}$ (with $M=1$~GeV as before) \cite{Graham:2020kai}, we find $\omega^\oplus_\vphi ({\rm proton}) \sim  3\times 10^{-3}$~rad/s.  Thus, the proton EDM storage ring proposal could probe our scenario if the long range scalar $\vphi$ couples  to proton spin.\footnote{We thank Y. Semertzidis for pointing out this possibility.}

In addition, similar measurements of electron $g-2$ can provide probes of $\vphi$ interaction with the electron spin, as a  different observable which depends on a new coupling $g_a^e$.  Since the above  scenario crucially depends on the scalar coupling $g_s^N$ to bulk matter, future improvements in bounds on long range forces in the relevant distance regime would also constrain our model.  

We close by noting that perhaps the most direct test of our scenario would be through the measurement of the muon $g-2$,  either deep inside the Earth or else far away from it, {\it e.g.} on the Moon.  Needless to say, either of those avenues would pose enormous technical and funding  challenges, the discussion of which lies  outside the scope of this Letter.     

\begin{acknowledgments}  We thank Y. Semertzidis for helpful comments on a draft version of this paper. The  work of H.D. and R.S. is supported by the US Department of Energy under Grant Contract DE-SC0012704.
\end{acknowledgments}
\vskip0.5cm

\appendix

{\it Appendix: relaxation of supernova constraints.}---Here, we argue that the supernova bound on $g_a^\mu$ obtained in Ref.~\cite{Bollig:2020xdr} can be avoided in the context of our model.  To see this, note that supernova bound relevant to the ``trapping regime" was based on the notion that surface emission of axions extended out less than 18~km, due to the paucity of muons beyond this radius.  This can be understood by noting that the temperature of the core at that radius is $T'\sim 10$~MeV, compared to the maximum value considered $T\sim 60$~MeV in Ref.~\cite{Bollig:2020xdr}.  Hence, the muon density at that radius is Boltzmann suppressed by 
\beq
\left(\frac{T'}{T}\right)^{(3/2)} e^{-\dfrac{m_\mu(T-T')}{T T'}}\sim \ord{10^{-5}}\,, 
\label{s}
\eeq
which is consistent with the results of Ref.~\cite{Bollig:2020xdr}.  The trapping surface for axions at radius $r_a$ has emission luminosity $L_a$ given by \cite{Bollig:2020xdr} 
\beq
L_a(r_a,T) = \frac{\pi^2}{120} (4 \pi r_a^2)T^4\,.
\label{La}
\eeq
We then find $L_a(18~\text{km}, 10~\text{MeV})\sim 10^{53}$~erg/s, which is in excess of the allowed energy transfer from the supernova core during the initial explosion, ruling out large values $g_a^\mu\gsim 10^{-6}$ that lead to trapped axions \cite{Bollig:2020xdr}.

The above analysis may not be applicable in our model, since $\vphi$ is allowed to have Yukawa couplings to SM fermions.  In particular, assuming an interaction $g_s^\mu \vphi \bar \mu \mu$, with $g_s^\mu\sim 10^{-18}$ which is easily allowed by constraints on long range forces  \cite{Davoudiasl:2018ltz} (see also Ref.~\cite{Carroll:2009dw} for related discussions), we argue that the trapping regime can again open up, and values of $g_a^\mu \lsim 10^{-4}$ adopted in our discussion in the main text can be allowed.  To see this, note that the conservative model chosen in Ref.~\cite{Bollig:2020xdr} for their final results corresponds $T\sim 30$~MeV in the central $\sim 10$~km of the supernova and a muon number density $\sim$ few $\times 10^{-3}$~fm$^{-3}$; we will adopt $n_\mu \sim 3 \times 10^{-3}$~fm$^{-3}$.  The field generated by muons in that region can shift the muon mass by 
\beq
\delta m_\mu = g_s^\mu \vphi \sim (g_s^\mu)^2 n_\mu\, r^2\,.
\label{delmmu}
\eeq
For the reference values above, we have $\delta m_\mu \gsim m_\mu$ which indicates that the presence of $\vphi$ in the theory can significantly affect the kinematics.  We note that in this regime of parameters, typically the plasma frequency $\omega_p \sim g_s^\mu (n_\mu/E)^{1/2}\gg m_\vphi$ and $\vphi$ may be considered massless. Then, assuming roughly 
constant $n_\mu$ over the region of interest, $\vphi\approx -m_\mu/g_s^\mu$ and hence the effective muon mass $m_{\rm eff}\equiv m_\mu + g_s^\mu \vphi$ would be driven to zero \cite{Domenech:2021uyx}.  

The preceding arguments suggest that the core of the supernova could have a much larger population of muons as $m_{\rm eff}/T\to 0$ and the trapping surface can extend well beyond what was considered in Ref.~\cite{Bollig:2020xdr}.  By making muons into effectively light degrees of freedom, they are no longer Boltzmann suppressed in outer parts of the supernova, with lower temperatures.  Therefore, one could potentially extend the trapping region to $\sim 100$~km, characterized by $T\sim 1$~MeV \cite{Sumiyoshi:2008qv}, where $L_a(100~\text{km}, 1~\text{MeV})\sim 10^{51}$~erg/s, which is no longer in conflict with axion emission bounds \cite{Bollig:2020xdr}.


\bibliography{earthly-Refs}

\end{document}